\documentstyle[12pt]{article}
\newcommand{\be}{\begin{equation}}
\newcommand{\ee}{\end{equation}}
\newcommand{\bea}{\begin{eqnarray}}
\newcommand{\eea}{\end{eqnarray}}
\newcommand{\ba}{\begin{array}}
\newcommand{\ea}{\end{array}}
\newcommand{\beas}{\begin{eqnarray*}}
\newcommand{\eeas}{\end{eqnarray*}}
\newcommand{\bes}{\begin{equation*}}
\newcommand{\ees}{\end{equation*}}

\begin{document}
\title{\bf Langevin diffusion in holographic backgrounds with hyperscaling violation}
\author{J. Sadeghi \thanks{Email:pouriya@ipm.ir}\hspace{1mm},  F. Pourasadollah \thanks{Email:f.pourasadollah@stu.umz.ac.ir}\\
{\small {\em  Sciences Faculty, Department of Physics, Mazandaran University,}}\\
{\small {\em P.O.Box 47416-95447, Babolsar, Iran}}}
 \maketitle
\begin{abstract}
 In this note we consider a relativistic heavy quark which moves in the quark-gluon plasmas. By using the holographic methods, we analyze the Langevin
diffusion process of this relativistic heavy quark. This heavy quark
is described by a trailing string attached to a
 flavor brane and moving at constant velocity. The fluctuations of this string are related to the thermal correlators
 and the correlation functions are precisely the kinds of objects that we compute in the gravity dual picture.
 We obtain the action of the trailing string in hyperscaling violation backgrounds and we then find
 the equations of motion. These equations lead us to construct the Langevin correlator which helps us to obtain the Langevin constants.
Using the Langevin correlators we derive the  densities spectral and
 simple analytic expressions in the small and large frequency limits. We examine our works for planar
 and $R$-charged black holes with hyperscaling violation and find new constraints on $\theta$ in the presence of velocity
 $v$.
\end{abstract}
\section{Introduction}
It is certainly important and interesting to understand about the
strongly coupled quark gluon plasma (QGP) [1-4], since heavy ion
collisions experiments has provided a variety of evidences for
creation of QGPs at RHIC. Over the recent years, there have been a
lot of efforts to study the features of heavy-ion collisions and the
QGP. In this context, the AdS/CFT correspondence [5-8] has provided
a powerful tool to study strongly coupled field theory. It maps
relativistic conformal field theories holographically to
gravitational (or stringy) dynamics in
a higher dimensional spacetime. This gauge/gravity duality provides the possibility of computing some properties of QGP.\\
QGPs can be thought of as a soup of quarks and gluons. A heavy quark immersed in this fluid, can be modeled in string
 theory (via the AdS/CFT correspondence) by an open string attached to the boundary of a bulk black hole.
 The end-point of this string receives to the heavy quark on
a boundary which is stretching in the UV part of bulk geometry. At a
classical level, the straight string is a solution to
 the equation of motion and dose not move in the absence of external force. In this case, the string extends from the
 boundary to the black hole horizon (at $r=r_{h}$). On the field theory side, a competition between
 the drag and the noise is balanced. And also the modes of string are in equilibrium
 at the Hawking temperature. The effect of thermal noise is not often considered in AdS/CFT. This seems to conflict
  with the fluctuation-dissipation theorem [9]. Clearly, one can except that the Hawking radiation, which is emitted
  from the black brane, persuade the string to have a random motion. The fluctuations caused by the Hawking radiation
  are integrated within the stretched horizon $r_{s}=r_{h}+\epsilon$.
  This gives a picture of the stochastic behavior of the string fluctuations as originating from the world-sheet horizon
    with the required noise at this horizon [10].
  The fluctuations of the trailing string (quantum) provide the information about the heavy
 quark as it moves in the plasma. So, the dynamics of fields on the boundary can be dictated by the
 effective action at the stretched horizon [11-12].\\
In this new scheme, in analogy with the dynamics of heavy quarks in
heat bath giving rise Brownian motion[13-17],
 one can consider the stochastic nature of the out of equilibrium systems. This involves a diffusive process,
 that was first considered by using the Schwinger-Keldysh formalism adapted to AdS/CFT [18].
  Another important improvement in this picture is related to relativistic Langevin evolution of
  the trailing string which is studied by [19-21]. The stochastic motion
  was formulated as a Langevin process [11-14] associated to the
  correlators of the fluctuations of the string.\\
On the practical perspective, one can consider a fundamental string whose end-point lies in the UV region
of a bulk black-hole background. The end-point of string is forced to move with velocity $v$.
The string stretches in the bulk until the stretched horizon, in a way that it becomes completely horizontal.
 When the quark is not moving, (or moving with $v\rightarrow0$) the stretched horizon approaches the black hole horizon.
 The classical profile of the trailing string can be obtained by solving the Nambu-Goto equation of motion.
  By considering small fluctuations around the classical string profile at the quadratic level, second-order radial equations are obtained.
  These fluctuations are related to the thermal correlators with modified temperature $T_{s}$, through the second-order radial equations,
  and satisfy the fluctuation-dissipation
 relation associated with this temperature. Since in this case the system is out of equilibrium, the Hawking temperature of the induced string
  world-sheet metric $T_{s}$ is in general different from the heat bath temperature $T$. A relativistic Langevin diffusion equations is
   associated with the correlators of the fluctuations of the string by the holographic prescription. The Langevin correlators obey the modified Einstein
 relation with modified temperature $T_{s}$ and the relation between the diffusion constants is changed with this modified Einstein relation.
   [22-23]. \\
There is already a huge amount of literature on the subject of holographic construction
of a heavy quark immersed in the quark gluon plasma. The motion of such a quark have been studied holographically
 in the classical and relativistical way in [14-17] and [22-23].
 In various articles the entries have been assigned to the investigation of sting fluctuations
  in the gravity theories where the corresponding plasmas have different features (e.g. rotation, charge and,...)
   [16-17], [24-25] and [30]. The construction of some holographic setups in the literature
  is formed so that the boundary theory is not conformally invariant.
  The holographic techniques have been used in the study of the submerging
quark in such plasmas in [22-23] and [30].
 The purpose of the present paper is to investigate the relativistic Langevin evolution
  of a heavy quark  in backgrounds with hyperscaling violation [26-33].\\
Hyperscaling is a feature of the free energy based on naive dimension. For the theories with hyperscaling, the entropy behaves
 as $ S\sim T^{d/z}$ where $T$, $d$ and $z$ are temperature, number of spatial dimensions and dynamical exponent respectively.
  Hyperscaling violation  first mentioned in context of holographic in Ref. [34]. In this context, the hyperscaling violation
  exponent $ \theta$ is related to the transformation of the proper distance, and its non-invariance implies the violation of
  hyperscaling of the dual field theory . Then, the relation between the entropy and temperature has been modified as
  $ S\sim T^{(d-\theta)/z}$. In general, theory with hyperscaling violation $d-\theta$ plays the role of an effective space
  dimensionality for the dual field theory. In theories with hyperscaling violation, the metric backgrounds are the sophisticated
  generalization of the AdS gravity. These metrics have a special characteristic so that they can be dual to the field theories which are not
  conformally invariant. The observations [26-27] and [35-36] indicate that backgrounds whose asymptotic
  behavior coincides with these metrics  may be of interest to condensed matter physics. So,   it  is natural to further
explore gauge/gravity duality for these backgrounds.\\
This paper is structured as follows: In section 2 we present the
description of backgrounds with
 hyperscaling violation and the relevant classical trailing string solution in these backgrounds. In section 3 we carry out the
 corresponding linear fluctuations, these fluctuations are utilized for the holographic computation of
the Langevin correlators.  also,  we obtain the Langevin
coefficients and the density spectral  associated to the Langevin
correlators. In sections 2 and 3 our computations are devoted to the
planar black holes with hyperscaling violation. But, in section 4 we
investigate the above mentioned  computations for $R$-charged black
hole with hyperscaling violation. In section 5, we summarize our
works  and make some comments for future reserach.
\section{Backgrounds with hyperscaling violation}
 In this section, we implement the backgrounds with hyperscaling violation as the bulk geometries.
 The string extends into the bulk is a dual of a heavy external quark moving through the plasma on the boundary of bulk.
 In order to discuss the Langevin coefficients of a heavy quark with gauge/gravity duality techniques,
 we have to find the fluctuations of the trailing string. These fluctuations are related to the thermal
 correlators and the correlation functions are precisely the kinds of objects that we compute
in the gravity dual picture. Therefore we obtain the action of the trailing string in the backgrounds
 with hyperscaling violation and then we find the equations of motion from that action. By solving these
  equations we are able to find the Langevin correlator which helps us obtain the Langevin constants.
 \subsection{Planar black holes with hyperscaling violation}
 As we indicated before, we want to utilize backgrounds with hyperscaling violation as the bulk geometries.
  These geometries arise generically as the solutions in appropriate Einstein-Maxwell-dilaton theories with the following
   action [33-34] and [39],
\begin{equation}
S=-\frac{1}{16\pi G} \int d^{d+2}x\sqrt{-g}\left[\Re-\frac{1}{2}(\partial\phi)^{2}-f(\phi)F_{\mu\nu}F^{\mu\nu}+V(\phi)\right],
\end{equation}
where $\phi$ is dilaton field and the symbols $g$ and $\Re$ are determinant of the metric and the scalar curvature respectively.
 The gauge coupling $g^{2} =(f(\phi))^{-1}$ and the potential $V (\phi)$ are both a function of the dilaton.
The black hole solution with hyperscaling violation from the action
(9) can be written as [26] and [39],
\begin{eqnarray}
d s^{2}_{d+2}&=& (\frac{R}{r})^{2}(\frac{r}{r_{F}})^{\frac{2\theta}{d}}\left[-r^{-2(z-1)}h(r)dt^{2}+h(r)^{-1}dr^{2}+
d x_{i}^{2}\right],\nonumber\\
h(r)&=&1-(\frac{r}{r_{h}})^{d+z-\theta},
\end{eqnarray}
We note here the metric background includes a dynamical critical exponent $z$ and a hyperscaling violation exponent $\theta$,
also $d$ is the number of transverse dimensions and  $i=1,...,d$,  $r=r_{h}$ is the location of horizon and $r=r_{F}$ is the boundary.
This metric is not scale invariant and under the following scaling,
\begin{equation}
t\rightarrow \lambda^{z}t  \qquad   , \quad x_{i}\rightarrow \lambda
x_{i} \quad ,\qquad r\rightarrow \lambda r.
\end{equation}
and transforms as,
\begin{equation}
ds\rightarrow \lambda ^{\frac{\theta}{d}}ds,
\end{equation}
which is defining property of hyperscaling in holographic language.
In order to understand the metric properties of this class of
spacetimes, notice that (2) is conformally equivalent to a Lifshitz
geometry [40-41] as can be seen after a Weyl rescaling
$g_{\mu\nu}\rightarrow \tilde{g}_{\mu\nu}=\Omega^{2}g_{\mu\nu}$,
with $\Omega=r^{-\frac{\theta}{d}}.$
The scale-invariant limit is $\theta = 0$, which reduces to a Lifshitz solution.\\
 It is reasonable from the gravity side to demand that the null energy condition (NEC) [26] and [33] be satisfied.
For metric background (2), this impose some constraints on $\theta$
and $z$ as,
\begin{eqnarray}
  (d-\theta)(d(z-1)-\theta)\geq 0,\qquad\qquad(z-1)(d+z-\theta)\geq
  0.
  \end{eqnarray}
These constraints have important consequences: First, in a Lorentz
invariant theory, $z = 1$ and then the first inequality implies that
$\theta \leq 0$ or $ \theta\geq d$. On the other hand, for a scale
invariant theory $(\theta = 0)$, we recover the known result $z\geq
1$. Notice that in theories with hyperscaling violation the NEC can
be satisfied for $z < 1$, while this range of dynamical exponents is
forbidden if $\theta = 0$. In particular, $\{z < 0; \theta > d\} $
gives a consistent solution to (5), as well as $\{0 < z < 1; \theta
\geq d + z\}$. The NEC gives $\theta > d$, but this range for
$\theta$ leads to instabilities in the gravity side. So this choice
of $\theta$ dose not lead to the physically consistent theories.
\subsection{Trailing string and drag force}
Before going in detail we should indicate that we review the calculation of the unperturbed trailing solution,
 that was discussed in [22-23]. We consider an external heavy quark which moves in the quark-gluon
 plasma medium with a fixed velocity $v$ on the boundary theory. It can be realized as the endpoint of an open classical trailing
 string which is hanged from the boundary and moves at constant velocity $v$. The dynamics of
this string is governed by the Nambu-Goto action,
\begin{equation}\
 S_{NG}=-\frac{1}{2\pi \acute{\alpha}}\int d\tau d\sigma
 \sqrt{-\det g_{_{ab}}}\,,
\end{equation}
where $g_{ab}= G_{\mu \nu}\partial_{a}X^{\mu}\partial_{b}X^{\nu}$ denotes the components of
 the bulk metric in the string frame. We choice the $\tau=t$ and $\sigma=r$ to work in static
  gauge and take the following anstaz for trailing string,
\begin{equation}
X^{1}=vt+\xi(r),\qquad\qquad X^{2},X^{3},...X^{d}=0,
\end{equation}
By using the metric background (2), the induced metric on the world-sheet can be obtained as,
\begin{equation}\label{eq 2}
g_{ab}= (\frac{R}{r})^{2}(\frac{r}{r_{F}})^{\frac{2\theta}{d}}
\left(
\begin{array}{ll}
v^{2}-r^{-2(z-1)}h(r)  & \hbox{$ v\xi'(r)$} \\
v\xi'(r)   & \hbox{$h(r)^{-1}+\xi'^{2}$}
\end{array}
\right).
\end{equation}
So, the corresponding action becomes,
\begin{equation}
 S_{NG}=-\frac{1}{2\pi \acute{\alpha}}\int dt dr (\frac{R}{r})^{2}(\frac{r}{r_{F}})^{\frac{2\theta}{d}}
 \sqrt{r^{-2(z-1)}-\frac{v^{2}}{h(r)}+r^{-2(z-1)}h(r)\xi'^{2}(r)}.
\end{equation}
The conjugate momentum $\pi_{\xi} $ flowed from the boundary to the
bulk and interpreted as the total force experienced by the quark, is
written by,
\begin{equation}
\xi'^{2}=\frac{\left(v^{2}-r^{-2(z-1)}h(r)\right)C^{2}}{\left[C^{2}-(\frac{R}{r})^{4}(\frac{r}{r_{F}})^{\frac{4\theta}{d}}r^{-2(z-1)}
h(r)\right]r^{-2(z-1)}h^{2}(r)},
\end{equation}
through the relation $C=2\pi \acute{\alpha}\pi_{\xi}$. , The horizon
for the induced world-sheet metric  is obtained,
\begin{equation}
r_{s}^{-2(z-1)}h(r_{s})=v^{2}.
\end{equation}
Since the numerator of the square root in (10) vanishes at the
stretched horizon, reality requires that the denominator also
vanishes. So, we
have$\pi_{\xi}^{2}=(\frac{R}{r_{s}})^{4}(\frac{r_{s}}{r_{F}})^{\frac{4\theta}{d}}r_{s}^{-2(z-1)}
\frac{h(r_{s})}{4\pi^{2} \acute{\alpha}^{2}}$. For $z=1$  and the
stretched horizon is given by,
\begin{equation}
r_{s}=r_{h}(1-v^{2})^{\frac{1}{d+1-\theta}}.
\end{equation}
In the special case $\theta=d+1$ the stretched horizon in (12) tends to the infinity. It is an unacceptable case,
since we except that the stretched horizon is smaller than the horizon $r_{h}$. Moreover,
 we eliminate $\theta>d+1$ for the similar reason.\\ The drag force on the quark can be obtained from
 the momentum conjugate which has a following form,
\begin{eqnarray}
F_{drag}&=&\pi_{\xi}=-\frac{vR^{2}r_{s}^{\frac{2\theta}{d}-2}}{2\pi \acute{\alpha}r_{F}^{2\frac{\theta}{d}}}\nonumber\\
F_{drag}&=&-\frac{R^{2}r_{h}^{2(\frac{\theta}{d}-1)}v(1-v^{2})^{\frac{2(\frac{\theta}{d}-1)}{d+1-\theta}}}
{2\pi \acute{\alpha}r_{F}^{2\frac{\theta}{d}}}\qquad\qquad \mathrm{for} \qquad z=1.
\end{eqnarray}
In the ultra-relativistic limit $v\rightarrow1$ the drag force for the case $z=1$ vanishes,
 but one can check that this event dose not happen for the range of $z>1$. The momentum friction coefficient $\eta$,
 responsible for the gradual loss in the momentum of a quark of mass
 $M$. It is related to the drag force via the $F_{drag}=-\eta p$, with $p=M\gamma v$ [42]. So, one can obtain,
\begin{equation}
\eta=\frac{R^{2}r_{h}^{2(\frac{\theta}{d}-1)}(1-v^{2})^{\frac{2(\frac{\theta}{d}-1)}{d+1-\theta}}}{2\pi
 \acute{\alpha}M\gamma r_{F}^{2\frac{\theta}{d}}}.
\end{equation}
where $\gamma=\frac{1}{\sqrt{1-v^{2}}}$ is the relativistic contraction factor. For the special case $\theta=d$,
 the drag force and momentum friction coefficient reduce to the
 following expression,
\begin{equation}
F_{drag}=-\frac{R^{2}v}{2\pi \acute{\alpha}r_{F}^{2}},\qquad\qquad
\eta=\frac{R^{2}}{2\pi \acute{\alpha}M\gamma r_{F}^{2}}.
\end{equation}
In this case, $F_{drag}$ and $\eta$ are independent of the black hole horizon.\\
If we diagonalize the induced world-sheet metric by transforming the coordinate trough the transformation
 $dt\rightarrow dt-\frac{v\xi'(r)}{ v^{2}-r^{-2(z-1)}h(r)}dr$, the resulting metric components are,
\begin{equation}
h_{tt}=(\frac{R}{r})^{2}(\frac{r}{r_{F}})^{\frac{2\theta}{d}}(-r^{-2(z-1)}h(r)+v^{2}),\qquad\qquad
 h_{rr}=\frac{r^{-2(z-1)}(\frac{R}{r})^{6}(\frac{r}{r_{F}})^{\frac{6\theta}{d}}}{r^{-2(z-1)}(\frac{R}{r})^{4}
(\frac{r}{r_{F}})^{\frac{4\theta}{d}}h(r)-C^{2}}.
\end{equation}
The effective Hawking temperature associated to the above black hole metric can be found,
\begin{equation}
T_{s}^{2}=\frac{1}{16\pi^{2}}\left[r_{s}^{-2(z-1)}h^{2}(r_{s})\left(\frac{h'(r_{s})}{h(r_{s})}+
\frac{(\frac{4\theta}{d}-2(z+1))}{r_{s}}\right)\left(\frac{h'(r_{s})}{h(r_{s})}+
\frac{-2(z-1)}{r_{s}}\right)\right].
\end{equation}
For $z=1$, the above relation reduces to,
\begin{equation}
T_{s}=\frac{1}{4\pi}\sqrt{\frac{\left[(\frac{4\theta}{d}+d-3-\theta)(1-v^{2})-
(\frac{4\theta}{d}-4)\right](1-v^{2})(d+1-\theta)}
{\left[r_{h}(1-v^{2})^{\frac{1}{d+1-\theta}}\right]^{2}}}.
\end{equation}
In order to have correct value for $T_{s}$ in the case of
$\theta<d+1$, we need to have following condition,
\begin{equation}
\theta<\frac{d\left[1+d+(3-d)v^{2}\right]}{d+(4-d)v^{2}}.
\end{equation}
In the special case $d=3$, we have,
\begin{equation}
\theta<\frac{12}{3+v^{2}}.
\end{equation}
The Hawking temperature related to the black hole horizon in the presence of
 hyperscaling parameter and dynamical exponent $z=1$ is defined as,
\begin{equation}
T=\frac{(d+1-\theta)}{4\pi r_{h}}.
\end{equation}
 From the equations (18) and (21), one can easily find that,
\begin{equation}
T_{s}^{2}=T^{2}\left[1-\frac{(\frac{4\theta}{d}+d-3-\theta)v^{2}}{(d+1-\theta)}\right](1-v^{2})^{1-\frac{2}{d+1-\theta}},
\end{equation}
For the special case $\theta=d$, it is obvious from the above
relation that the Hawking temperature and the modified temperature
become equal. This equality is also confirmed for the range $z>1$.
In the conformal limit, it means the hyperscaling parameter tends to
zero and the background solution reduces to AdS-Schwarzschild. So,
 the stretched horizon position and temperature are given by,
 \begin{equation}
T_{s}^{2}=T^{2}[1-\frac{d-3}{d+1}v^{2}](1-v^{2})^{1-\frac{2}{d+1}},\qquad
r_{s}=\frac{d+1}{4\pi T}(1-v^{2})^{\frac{1}{d+1}}.
\end{equation}
For $d=3$ we receive to the expected relation [23],
 \begin{equation}
T_{s}=\frac{T}{\sqrt{\gamma}},\qquad r_{s}=\frac{1}{4\pi \sqrt{\gamma}T}
\end{equation}
\subsection{Fluctuations of the Trailing String}
In order to study the stochastic motion of  quark, we proceed to
investigate the fluctuations around the classical trailing solution.
 We choose the static gauge, such that the string embedding becomes $X^{\mu}(t,r)=(t,r,X^{1}(t,r),X^{2}(t,r),...,X^{d}(t,r)$.
 So, we take the following anstaz for embedding,
\begin{equation}
X^{1}(t,r)=vt+\xi(r)+\delta X^{\|}(t,r)\quad,\quad X^{2}(t,r)=\delta X^{2}(t,r)\quad,...\quad,X^{d}(t,r)=\delta X^{d}(t,r).
\end{equation}
By expanding the Nambu-Goto action in $ \delta X^{i}(t,r)$ around the classical solution up to quadratic terms we have,
\begin{equation}
 S_{2}=-\frac{1}{2\pi\acute{\alpha}}\int dt dr \frac{H^{ab}}{2}\left[N(r)\partial_{a}\delta X^{\|}\partial_{b}\delta X^{\|}+
 \sum_{i=2}^{d}G_{ii}\partial_{a}\delta X^{i}\partial_{b}\delta X^{i}\right]
\end{equation}
where,
\begin{equation}
 N(r)= \frac{(r^{-2(z-1)}h(r)\frac{R}{r})^{4}(\frac{r}{r_{F}})^{\frac{4\theta}{d}}-C^{2}}
 {(\frac{R}{r})^{2}(\frac{r}{r_{F}})^{\frac{2\theta}{d}}(r^{-2(z-1)}h(r)-v^{2})},\qquad\qquad G_{ii}=(\frac{R}{r})^{4}
 (\frac{r}{r_{F}})^{\frac{4\theta}{d}}
\end{equation}
and $H^{ab}=\sqrt{-h}h^{ab}$. The equations of motion can be found from the above action,
\begin{equation}
\partial_{a}(H^{ab}N(r)\partial_{b}\delta X^{\parallel})=0,\qquad\quad \partial_{a}(H^{ab}G_{ij}\partial_{b}\delta X^{\perp})=0,
\end{equation}
The definitions of $\delta X^{\perp}$ and $\delta X^{\parallel}$ are
responsible for the longitudinal and the transverse fluctuations
respectively. By taking a harmonic anstaz as $\delta
X^{i}(r,t)=e^{i\omega t}\delta X^{i}(r,\omega)$, the equations (28)
become,
\begin{eqnarray}
 &&\partial_{r}\left[\frac{\sqrt{\left(C^{2}-(\frac{R}{r})^{4}
(\frac{r}{r_{F}})^{\frac{4\theta}{d}}r^{-2(z-1)}h(r)\right)\left(v^{2}-r^{-2(z-1)}h(r)\right)
}}{r^{-(z-1)}}\partial_{r}\delta X^{\perp}\right]\nonumber\\
&&+\frac{\omega^{2}r^{-(z-1)}(\frac{R}{r})^{4}
 (\frac{r}{r_{F}})^{\frac{4\theta}{d}}}{\sqrt{\left(C^{2}-
(\frac{R}{r})^{4}
 (\frac{r}{r_{F}})^{\frac{4\theta}{d}}r^{-2(z-1)}h(r)\right)\left(v^{2}-r^{-2(z-1)}h(r)\right)}}\delta X^{\perp}=0
\end{eqnarray}
\begin{eqnarray}
 &&\partial_{r}\left[\frac{\left(C^{2}-(\frac{R}{r})^{4}
 (\frac{r}{r_{F}})^{\frac{4\theta}{d}}r^{-2(z-1)}h(r)\right)^{\frac{3}{2}}}{r^{-(z-1)}\left(v^{2}-r^{-2(z-1)}
 h(r)\right)^{\frac{1}{2}}}\partial_{r}\delta X^{\parallel}\right]\nonumber\\
&&+\frac{\omega^{2}r^{-(z-1)}\left(C^{2}-
(\frac{R}{r})^{4}(\frac{r}{r_{F}})^{\frac{4\theta}{d}}r^{-2(z-1)}h(r)\right)^{\frac{1}{2}}}{\left(v^{2}-
r^{-2(z-1)}h(r)\right)^{\frac{3}{2}}}\delta X^{\parallel}=0
\end{eqnarray}
In what follows, we will construct solutions to equations (29) and
(30) for the string fluctuations and obtain the diffusion constants
and the density spectral  from them. However, by using the method of
the membrane paradigm, computation of diffusion constants can be
done directly from the quadratic action (26). We will derive these
constants through the two different methods in the next section.
\section{Holographic computation of Langevin correlators and diffusion constants }
 \subsection{Momentum correlators from the trailing string}
The Langevin correlators can be computed holographically from the classical solutions for the fluctuations of the trailing string.
 Two types of independent retarded correlators $\textit{g}_{R}^{\parallel}$ and
  $\textit{g}_{R}^{\perp}$ for the longitudinal and transverse fluctuations
  [19] are reasonably expected from the structure of the action (26) for the fluctuations,
\begin{equation}
\textit{g}_{\parallel}^{ab}=\frac{N}{2\pi\acute{\alpha}}H^{ab},\qquad\qquad \textit{g}_{\perp}^{ab}=
\frac{G_{ii}}{2\pi\acute{\alpha}}H^{ab}.
\end{equation}
In the holographic version for the retarded correlator of diagonal metric (16) we have,
\begin{equation}
G_{R}(\omega)=-[\psi^{*}(r,\omega)\textit{g}^{rr}\partial_{r}\psi(r,\omega)]_{boundary},
\end{equation}
where $\psi$ is related to the fluctuations $\delta X^{\parallel}$ and $\delta X^{\perp}$.
The expression in equation (32) must be evaluated at the boundary of the trailing string world-sheet.
In theories with hyperscaling violation, existence of a dimensionful scale that dose not decouples in
the infrared, requires the proper powers of this scale, which is denoted by $r_{F}$.
 Also we note here, by following the effective holographic approach [43] in which the dual theory lives on a finite $r$ slice,
 the metric background (2) provides a good description of the dual field theory only for a certain range of r, presumably
  for $r\geq r_{F}$ anticipating the applications at the low energy regions. In the case of an infinitely massive quark,
  the string is attached at the AdS boundary at $r=0$, however in our case with hyperscaling violation,
  the string is connected to the boundary at $r=r_{F}$ where $r_{F}\rightarrow 0$. In the case of finite mass quark,
  the trailing string is attached to a point $r_{b}$ and its mass is given through the following relation,
\begin{equation}
M=\frac{1}{2\pi\acute{\alpha}}\int_{0}^{r_{b}}(\frac{R}{r_{F}^{\frac{\theta}{d}}})^{2}r^{\frac{2\theta}{d}-2-z+1}=
\frac{1}{2\pi\acute{\alpha}}(\frac{R}{r_{F}^{\frac{\theta}{d}}})^{2}\frac{r_{b}^{\frac{2\theta}{d}-z}}{\frac{2\theta}{d}-z}.
\end{equation}
For $\theta\sim\frac{z}{2}d $, the mass of quark tends to the infinity. Moreover,
we expect an infinitely massive quark for $r_{b}\rightarrow0 $
. Thus, $\theta<\frac{z}{2}d$ is the acceptable region for
$\theta$ to make the expected mass for quark in the limit $r_{b}\rightarrow0$ [44].\\
The solutions of fluctuation equations (29) and (30) have a same
behavior for the transverse and longitudinal components at the
world-sheet
 horizon and at the boundary. At the $r\rightarrow r_{s}$ limit both equations take following form,
\begin{equation}
\partial_{r}^{2}\psi+\frac{1}{|r-r_{s}|}\partial_{r}\psi+(\frac{\omega}{4\pi T_{s}|r-r_{s}|})^{2}\psi=0,
\end{equation}
so, the solution in  near  world-sheet horizon is given by,
\begin{equation}
\psi(r,\omega)\sim (r_{s}-r)^{\pm i\frac{\omega}{4\pi T_{s}}},
\end{equation}
where $T_{s}$ is the modified temperature which is given by (17) and (18) for special case $(z=1)$.
In the above relation the outgoing waves are brought in $+$ sign while the incoming waves are in $-$ sign.\\
Near the boundary $r\rightarrow r_{F}$, for equations (29) and (30),
we have to discuss about the range of $\theta$ and $z$. Let us
consider $\theta<d+z$, where $h(r)$ vanishes in the limit of
$r\rightarrow r_{F}\rightarrow 0$. In the special case $z=1$ both
equations reduce to,
\begin{equation}
\partial_{r}^{2}\psi+\frac{\frac{2\theta}{d}-2}{r}\partial_{r}\psi+\omega^{2}\gamma^{2}\psi=0,
\end{equation}
therefore the solution of   above equation is given by,
\begin{equation}
\psi=c_{1}r^{\frac{3d-2\theta}{2d}}J_{-\frac{3d-2\theta}{2d}}(\gamma\omega r)+c_{2}r^{\frac{3d-2\theta}{2d}}
Y_{-\frac{3d-2\theta}{2d}}(\gamma\omega r),
\end{equation}
where for small $\omega$ two independent solutions with normalizable
and non-normalizable modes become,
\begin{equation}
\psi=c_{s}+c_{v}r^{\frac{3d-2\theta}{d}}.\\
\end{equation}
Near the boundary for the range of $z>0$, we have,
\begin{eqnarray}
  \textbf{U}&=&C^{2}-(\frac{R}{r})^{4}
 (\frac{r}{r_{F}})^{\frac{4\theta}{d}}r^{-2(z-1)}h(r)\rightarrow -\frac{R^{4}}{r_{F}^{2(z+1)}} \nonumber\\
  \textbf{V} &=&v^{2}-r^{-2(z-1)}h(r)\rightarrow v^{2} \qquad for \qquad z<1,\nonumber\\
 & \rightarrow& -r_{F}^{-2(z-1)} \qquad for \qquad z>1,
     \end{eqnarray}
Then,  the fluctuation equations (29) and (30) for $z>1$ near the
boundary for both cases  become,
\begin{equation}
\partial_{r}^{2}\psi+\frac{\frac{2\theta}{d}-1-z}{r}\partial_{r}\psi+\omega^{2}r^{2(z-1)}\psi=0,
\end{equation}
which has the following solutions,
\begin{equation}
\psi=c_{1}r^{\frac{(z+2)d-2\theta}{2d}}J_{-\frac{(z+2)d-2\theta}{2zd}}(\frac{\gamma\omega}{z} r^{z})+
c_{2}r^{\frac{(z+2)d-2\theta}{2d}}Y_{-\frac{(z+2)d-2\theta}{2zd}}(\frac{\gamma\omega}{z} r^{z}),
\end{equation}
Also, in the region $\omega r\ll1$ we receive following equation,
\begin{equation}
\psi=c_{s}+c_{v}r^{\frac{2d+zd-2\theta}{d}}.\\
\end{equation}
for $z<1$ these equations reduce to,
\begin{equation}
\partial_{r}^{2}\psi+\frac{\frac{2\theta}{d}-2}{r}\partial_{r}\psi-\frac{\gamma^{2}}{v^{2}}\psi=0,
\end{equation}
and the solution of  above equation will be as,
\begin{equation}
\psi=c_{1}r^{\frac{3d-2\theta}{2d}}I_{-\frac{3d-2\theta}{2d}}(\frac{\omega}{v}
r)+c_{2}r^{\frac{3d-2\theta}{2d}}K_{-\frac{3d-2\theta}{2d}}(\frac{\omega}{v}r),
\end{equation}
and for the small $\omega$ we receive to the relation (38).\\
The appropriate boundary conditions for the wave functions in the
expression (35) for the retarded correlator are in-falling behavior
at the world-sheet horizon with the condition $\psi (r)=1$ [45]:
\begin{eqnarray}
\psi(r_{b},\omega) =1\qquad\qquad r_{b}=\left\{
\begin{array}{ll}
 0 &  M\rightarrow\infty\\
r_{q}   &   M\quad finite\\
 \end{array}
\right.
\end{eqnarray}
\begin{equation}
\psi(r,\omega)\sim \psi_{h}(r_{s}-r)^{-i\frac{\omega}{4\pi T_{s}}}\qquad\qquad r\sim r_{s}.
\end{equation}
By utilizing the wave functions, we can construct the propagator from relation (32).
We consider the properties of real and imaginary parts of retarded
Green's functions separately.
\paragraph{Real part of retarded correlator}
In the real part of retarded correlator (32) there are some ambiguities related to the UV divergencies in the on-shell action.
To avoid these divergencies and receive to finite results, one has to investigate the action on a
regularized space-time with boundary at $r=r_{F}$ . Then, after identifying the
divergencies in the limit $r_{F}\rightarrow 0$, the counterterms can be added to prevent infinite results.\\
 To evaluate the real part of (32), the wave-functions close to the boundary can be implemented for different
 regions of $z$. For $z=1$, we can expand the solution (37) for $\gamma\omega r\ll1$ near the boundary $r=r_{F}$ as,
\begin{eqnarray}
\psi_{UV}(r)=c_{1}(\frac{\gamma\omega}{2})^{\frac{-3d+2\theta}{2d}}\left[\frac{1}{\Gamma(\frac{-1}{2}
+\frac{\theta}{d})}-\frac{(\gamma\omega)^{2}r_{F}^{2}}
{4\Gamma(\frac{1}{2}+\frac{\theta}{d})}\right]\qquad\qquad\qquad\nonumber\\
-\frac{c_{2}}{\pi}(\frac{\gamma\omega}{2})^{\frac{3d-2\theta}{2d}}r_{F}^{\frac{3d-2\theta}{d}}
\left[\Gamma(\frac{-3}{2}+\frac{\theta}{d})
+\frac{\Gamma(\frac{-5}{2}+\frac{\theta}{d})(\gamma\omega)^{2}r_{F}^{2}}{4}\right]+...,
\end{eqnarray}
According to the relation (45) the value of $c_{1}=c_{s}$  is fixed
at $r=r_{F}$,
\begin{equation}
c_{1}=c_{s}=(\frac{\gamma\omega}{2})^{\frac{3d-2\theta}{2d}}\Gamma(\frac{-1}{2}+\frac{\theta}{d}).
\end{equation}
To obtain the real part of (32) we must evaluate the relation (31) at the boundary, then we have,
\begin{equation}
H^{rr}\sim\frac{1}{\gamma},\qquad\qquad H^{tt}\sim -\gamma ,\qquad\qquad
N\sim \frac{\gamma^{2} R^{2}}{ r_{F}^{2}}.
\end{equation}
Eventually, we find the following divergent term from the expression (47) for the transverse and longitudinal components:
\begin{equation}
Re\,G_{R}^{\perp}\simeq \gamma^{-2} Re\,G_{R}^{\parallel}\simeq \frac{\gamma\omega^{2}}{2\pi\acute{\alpha}
(1-\frac{2\theta}{d})r_{F}}\left[
1+\frac{\gamma^{2}\omega^{2}r_{F}^{2}}{2(1-\frac{2\theta}{d})}+O(r_{F}^{4})\right].
\end{equation}
Notice that in derivation of the above relation we neglect the
effect of second term in equation (47), which is proportional to
$c_{2}=c_{v}$ and starts at
$O(r_{F}^{3-\frac{2\theta}{d}}\omega^{\frac{3d-2\theta}{2d}})$
(considering the region of $\theta<d+1$).
 The correlators have UV divergences (in the $\omega^{2}$ term) which arise from the scheme dependence in their
calculation.\\
We now address the analysis of on-shell action to obtain the transverse and longitudinal Green's functions and
 compare them with the results of relation (50). We study the divergence structure of the action (6), expanded to
  quadratic order in the fluctuations defined in equation (26), around the classical
trailing string solution [23]. So we write,
\begin{equation}
S_{NG}=S_{0}+S_{1}+S_{2}+...\:.
\end{equation}
For each term in the above relation we derive the divergency around $r=r_{F}$ separately. The zeroth order term reads simply:
\begin{equation}
S_{0}=\frac{-1}{2\pi \acute{\alpha}}\int dt dr\sqrt{-g}=\frac{-1}{2\pi \acute{\alpha}}
\int dt\int_{r_{F}}^{r_{s}}dr \frac{R^{2}}{\gamma r_{F}^{\frac{2\theta}{d}}}r^{\frac{2\theta}{d}-2}.
\end{equation}
Around $r=r_{F}$, the above integral shows a divergency of the order $\frac{1}{r_{F}}$ ,
\begin{equation}
S_{0}^{div}=\frac{R^{2}}{2\pi \acute{\alpha}(1-\frac{2\theta}{d})\gamma r_{F}}\int dt.
\end{equation}
For the second term in (31) the quadratic order action (26) is implemented. By inserting the solution (38)
 close to the boundary and using relation (49), we obtain the divergent part of (26) as,
\begin{eqnarray}
S_{2}^{div}&=&\frac{R^{2}}{2\pi \acute{\alpha}(1-\frac{2\theta}{d})\gamma r_{F}}\frac{1}{2}
\int d\omega \omega^{2}\left(\gamma^{2}|c_{s}^{\perp}(\omega)|^{2}+\gamma^{4}
|c_{s}^{\parallel}(\omega)|^{2}\right) \nonumber\\
&=&\frac{R^{2}}{2\pi \acute{\alpha}(1-\frac{2\theta}{d})\gamma r_{F}}\frac{1}{2}\int dt \gamma^{2}\left(\delta\dot{X}^{\perp}\right)^{2}+
\gamma^{4}\left(\delta\dot{X}^{\parallel}\right)^{2}.
\end{eqnarray}
One can easily check that there is no divergency coming from the first order action.
Consequently, from the above action the divergent parts of the
transverse and longitudinal Green's functions are,
\begin{equation}
\left(Re\,G_{R}^{\perp}\right)=\frac{R^{2}}{2\pi \acute{\alpha}(1-\frac{2\theta}{d}) r_{F}}\gamma\omega^{2},\qquad\qquad
\left(Re\,G_{R}^{\parallel}\right)=\frac{R^{2}}{2\pi \acute{\alpha}(1-\frac{2\theta}{d}) r_{F}} \gamma^{3}\omega^{2}.
\end{equation}
The resorption of both (53) and (54) divergencies can be done by adding a single covariant boundary counterterm,
\begin{equation}
S_{count}=\Delta M(r_{F})\int dt\sqrt{\dot{X}^{\mu}\dot{X}_{\mu}},
\end{equation}
where is responsible for the renormalization of the quark mass. By expanding the above relation to second
order in $\overrightarrow{X}=\overrightarrow{v}t+\delta \overrightarrow{X}$, we find:
\begin{equation}
S_{count}=\frac{\Delta M(r_{F})}{\gamma}\left\{\int dt+\int\frac{1}{2}\int dt \gamma^{2}\left(\delta\dot{X}^{\perp}\right)^{2}+
\gamma^{4}\left(\delta\dot{X}^{\parallel}\right)^{2}\right\}.
\end{equation}
It is clear from (53) and (54) that the following choice for $\Delta M$ eliminates the total divergencies:
\begin{equation}
\Delta M(r_{F})=-\frac{R^{2}}{2\pi \acute{\alpha}(1-\frac{2\theta}{d}) r_{F}}.
\end{equation}
If we repeat the above progress for the case of $z>1$ for the zeroth order we have,
\begin{equation}
S_{0}^{div}=\frac{R^{2}}{2\pi \acute{\alpha}(z-\frac{2\theta}{d}) r_{F}^{z}}\int dt.
\end{equation}
The divergency coming from the second order action is given by,
\begin{eqnarray}
S_{2}^{div}&=&\frac{R^{2}}{2\pi \acute{\alpha}(2-z-\frac{2\theta}{d}) r_{F}^{2-z}}
\frac{1}{2}\int d\omega \omega^{2}\left(|c_{s}^{\perp}(\omega)|^{2}+
|c_{s}^{\parallel}(\omega)|^{2}\right) \nonumber\\
&=&\frac{R^{2}}{2\pi \acute{\alpha}(2-z-\frac{2\theta}{d}) r_{F}^{2-z}}\frac{1}{2}\int dt
\left(\delta\dot{X}^{\perp}\right)^{2}+
\left(\delta\dot{X}^{\parallel}\right)^{2}.
\end{eqnarray}
So the divergent parts of the transverse and longitudinal Green's functions are identical as,
\begin{equation}
\left(Re\,G_{R}^{\perp}\right)=\left(Re\,G_{R}^{\parallel}\right)=\frac{R^{2}}{2\pi \acute{\alpha}
(2-z-\frac{2\theta}{d}) r_{F}^{2-z}}\omega^{2},
\end{equation}
where this result can also explicitly derived from the explicit expansions of the wave-functions (41) close to the boundary.
We note that to get the result of (60), we use the following relation,
\begin{equation}
H^{rr}\sim r_{F}^{-(z-1)},\qquad\qquad H^{tt}\sim -r_{F}^{z-1}  ,\qquad\qquad
N\sim \frac{\gamma^{2} R^{2}}{ r_{F}^{2}}.
\end{equation}
 By comparing two divergencies in the zeroth and the second order,
  we find that the zeroth order divergency dominate over the second order since $z>1$.
  For second order, the divergencies appear only in the range of $ 1<z<2 $.
  So we may consider the following term for renormalizing the quark mass and fixing the coefficient of the counterterm action,
\begin{equation}
\Delta M(r_{F})=-\frac{R^{2}\gamma}{2\pi \acute{\alpha}(z-\frac{2\theta}{d}) r_{F}^{z}}.
\end{equation}
Unfortunately, this choice for changes in the quark mass does not completely cancels divergencies coming from the on-shell action.
 It seems that the another counterterms should be added before to cancel these divergencies. However, we showed that
  the real part of retarded correlator are equal to divergencies which come from  the unrenormalized on-shell action.
  The discussion on removing the divergencies remains as a open problem we may investigate in the future.\\
In the region of $z<1$ all computation are similar to the case of $z=1$ except that in this region we must replace $\gamma$ with $\frac{1}{\sqrt{-v^{2}}}$.
 This means that in the renormalization of quark mass we have to add a virtual mass to the quark to
 receive the finite results and also we encountered to an imaginary value in the computation of the real part of retarded correlator.
 This seems somewhat complicated, but as we know this range of dynamical exponents is forbidden if $\theta=0$ [46].
 In particular, as discussed before the range $0 < z < 1$ with $\theta \geq d + z$
  gives a consistent solution to the Null Energy Condition, as well as for  $z < 0, \theta >0$.
  So one can conclude that the range  of $0 < z < 1$  is not in agreement with $\theta < d + z$
  (the range of $\theta$ that we assumed at the first of our work). In continue paper, we don't consider this region for $z$.
\paragraph{Imaginary part of retarded correlator.}
In the imaginary part of retarded correlator we don't encounter the divergencies,
 since it is proportional to the conserved quantity (current) as,
\begin{equation}
\mathrm{Im} \textit{G}_{R}(\omega)= -\frac{1}{2i} \emph{g} ^{rr}\psi_{R}^{*}\partial_{r}\psi_{R}\equiv-J^{r},
\end{equation}
and compute at the horizon. From the definitions (27) and (31), we
find in the near-horizon limit,
\begin{eqnarray}
&&\emph{g} ^{rr}_{\perp}\simeq 4\pi T_{s}(r_{s}-r)G_{ii}(r_{s}),
\qquad \emph{g} ^{rr}_{\parallel}\simeq 4\pi T_{s}(r_{s}-r)N(r_{s}),\quad
\quad r\rightarrow r_{s}, \nonumber\\
&&\mathrm{where}\qquad \textrm{N}(r_{s})=\frac{16
\pi^{2}R^{2}T_{s}^{2}}{r_{F}^{\frac{2\theta}{d}}h^{2}(r_{s})
\left(\frac{h'(r_{s})}{h(r_{s})}+
\frac{-2(z-1)}{r_{s}}\right)^{2}}r_{s}^{\frac{2\theta}{d}+2z-4}.
\end{eqnarray}
By substituting the above relation in  (64) and utilizing the
solution (46) for $\psi$  one can write,
\begin{equation}
\mathrm{Im} \textit{G}_{R}^{\perp}=-\frac{\textit{G}_{ii}(r_{s})\mid
\psi_{h}^{\perp}\mid^{2}}{2\pi \acute{\alpha}}\omega
,\qquad\qquad\mathrm{Im}
\textit{G}_{R}^{\parallel}=-\frac{\textit{N}(r_{s}) \mid
\psi_{h}^{\parallel}\mid^{2}}{2\pi \acute{\alpha}}\omega\,
\end{equation}
From [11] and [47] we get the imaginary part of the retarded
correlator which is given by,
\begin{equation}
G(\omega) =\cot (\omega/(2T_{s}))\mathrm{Im}G_{R}(\omega).
\end{equation}
where $G_{R}(\omega)$ is symmetrized correlator.  The  density
spectral  associated to  Langevin dynamics are defined as,
\begin{equation}
\rho_{R}(\omega)=\frac{-1}{\pi}\mathrm{Im}G_{R}(\omega)\qquad\mathrm{and}\qquad
\rho(\omega)= \frac{-1}{\pi}\mathrm{Im}G(\omega)= \cot
(\omega/(2T_{s})) \rho_{R}(\omega).
\end{equation}
 The above equations help us to investigate the density spectral at the large frequency.
  In order to study the behavior of Longevin correlators
 at the high frequency we have to use the $WKB$ method.
\subsection{The WKB approximation at large frequency}
In this section we are going  to  derive the large-frequency limit
of the density spectral , so in order to do this process we need to
apply the WKB method. So, in order to arrange the equation as a
Schr\"{o}dinger-like form (29) and (30) equations,  we  have to
rescale the corresponding wave function. The large $\omega$ solution
can be obtained by an adaptation of the $WKB$ method [23],[48]. The
Schr\"{o}dinger-like form of this equations is,
\begin{equation}
-\phi''+V_{s}(r)\phi=0,\qquad\qquad
V_{s}(r)=-\frac{\omega^{2}G_{ii}^{2}}{L^{2}}+\frac{1}{2}\left(\log
\mathcal{L}\right)''+ \frac{1}{4}\left(\log \mathcal{L}\right)'^{2}.
\end{equation}
where  $\phi=\sqrt{\mathcal{L}}\psi$ with,
\begin{eqnarray}
\psi =\left\{
\begin{array}{ll}
\delta X^{\perp}  &  \,\\
\delta X^{\parallel}  &  \,\\
\end{array}
\right.   \,, \mathcal{L}=\left\{  \begin{array}{ll}
L &  \,\\
\frac{LN}{G_{ii}}   &  \,\\
\end{array}
\right. \qquad,\mathrm{L}=\frac{\sqrt{\left(C^{2}-(\frac{R}{r})^{4}
(\frac{r}{r_{F}})^{\frac{4\theta}{d}}r^{-2(z-1)}h(r)\right)
\left(v^{2}-r^{-2(z-1)}h(r)\right)}}{r^{-(z-1)}}.
\end{eqnarray}
In order to solve Schr\"{o}dinger equation with some approximations
, we divide the range $r_{F}<r<r_{H}$ in three region $r\ll r_{H}$,
$r\simeq r_{H} $ and $r _{tp}\ll r< r_{H}.$
\paragraph {Near Boundary}: $r\ll r_{s}$\\
 In this region for $z=1$ and $z>1$ we have different asymptotic as:
\begin{eqnarray}
L =\left\{
\begin{array}{ll}
\frac{R^{2}r^{\frac{2\theta}{d}-2}}{r_{F}^{\frac{2\theta}{d}}\gamma} & \,\\
\frac{R^{2}r^{\frac{2\theta}{d}-1-z}}{r_{F}^{\frac{2\theta}{d}}} &\, \\
\end{array}
\right.   \,, N=\left\{  \begin{array}{ll}
\frac{R^{2}r^{\frac{2\theta}{d}-2}}{r_{F}^{\frac{2\theta}{d}}}& \qquad z=1\\
\frac{R^{2}r^{\frac{2\theta}{d}-2}}{r_{F}^{\frac{2\theta}{d}}}  &\qquad z>1\\
\end{array}
\right.
\end{eqnarray}
By using these relations the Schr\"{o}dinger potential is given by,
\begin{eqnarray}
V_{s}\simeq\left\{
\begin{array}{ll}
-\gamma^{2}\omega^{2}+\frac{(\frac{\theta}{d}-1)(\frac{\theta}{d}-2)}{r^{2}} & z=1 \nonumber\\
\frac{-\omega^{2}}{r^{-2(z-1)}}+\frac{(\frac{\theta}{d}-\frac{1}{2}(z+1))(\frac{\theta}{d}-\frac{1}{2}(z+3))}{r^{2}}&
z>1 \nonumber\\
\end{array}
\right.
\end{eqnarray}
We replace these potential to the relation (69), then the solution
of Schr\"{o}dinger equation will be as,
\begin{equation}
\phi=A_{1}\sqrt{r}J_{\frac{3d-2\theta}{2d}}(r\gamma \omega)+A_{2}\sqrt{r}Y_{\frac{3d-2\theta}{2d}}(r\gamma \omega)
,\;\qquad\qquad\qquad z=1
\end{equation}
\begin{equation}
\phi=A_{1}\sqrt{r}J_{\frac{(z+2)d-2\theta}{2zd}}(\frac{r^{z}\omega}{z})+A_{2}\sqrt{r}Y_{\frac{(z+2)d-
2\theta}{2zd}}(\frac{r^{z}\omega}{z}),
\qquad\qquad z>1
\end{equation}
\paragraph{Near Horizon}: $r\simeq r_{s}$ \\
In this region for both $z=1$ and $z>1$ case we obtain,
\begin{equation}
L=(4\pi T_{s})\frac{R^{2}r_{s}^{\frac{2\theta}{d}-2}}{r_{F}^{\frac{2\theta}{d}}} (r_{s}-r).
\end{equation}
If we implement the above relation and equation (65) for $N(r_{s})$
in the schr\"{o}dinger potential one can arrive at,
\begin{equation}
V_{s}\simeq -(\tilde{\omega}^{2}+\frac{1}{4})\frac{1}{(r-r_{s})^{2}},\qquad\qquad r\rightarrow r_{s}
\end{equation}
where $ \tilde{\omega}=\frac{\omega}{4\pi T_{s}}$. The solution of
Schr\"{o}dinger equation (69) after substituting this potential
in-falling boundary condition at the horizon is given by,
\begin{equation}
\phi_{h}\simeq C_{h}\left(r_{s}-r\right)^{-i\tilde{\omega}+\frac{1}{2}}.
\end{equation}
\paragraph{WKB region}: $r_{tp}<r\ll r_{s}$\\
This region is allowed classically and it covers almost all  range
as $r_{F}<r<r_{s}$. For large $\omega$'s, the first term of equation
(69) dominates and the potential becomes,
\begin{equation}
V_{s}\simeq -\frac{\omega^{2}G_{ii}^{2}}{R^{2}},\qquad\qquad r_{tp}<r\ll r_{s}.
\end{equation}
For a small region close to the boundary, including turning point $r_{tp}$, the approximation (77) breaks down.
The turning point for large $\omega$'s is found by solving the equation $V_{s}(r) = 0$,
\begin{eqnarray}
r_{tp}=\left\{
\begin{array}{ll}
\frac{ \sqrt{(\frac{\theta}{d}-1)(\frac{\theta}{d}-2)}}{\gamma \omega} & z=1 \nonumber\\
\left[\frac{(\frac{\theta}{d}-\frac{1}{2}(z+1))(\frac{\theta}{d}-\frac{1}{2}(z+3))}
{\omega^{2}}\right]^{\frac{1}{2z}}&  z>1 \\
\end{array}
\right.
\end{eqnarray}
The crucial fact is that, for large $\omega$s, $r_{tp}\ll r_{s}$ the
regions 1 and 3 overlap. On the other hand, also regions 2 and 3
overlap and close to $r \simeq r_{s}$. Therefore, the solution in
WKB region can be
 used to connect the near-boundary and near-horizon asymptotic. By inserting the expression (77) in
 the equation (69)  two independent solutions to $ -\phi'' +V\phi = 0$ in the region $V \ll0$ are written as,
\begin{equation}
\phi_{1}=\frac{1}{\sqrt{p}}\cos\int^{r}p,\qquad \phi_{2}=\frac{1}{\sqrt{p}}\sin\int^{r}p,\qquad p(r)=\sqrt{-V_{s}(r)}.
\end{equation}
Explicitly, the general solution has the following form,
\begin{equation}
\phi_{wkb}=C_{1}\frac{\sqrt{L}}{G_{ii}}\cos\int^{r}\frac{\omega^{2}G_{ii}}{L}+C_{2}\frac{\sqrt{L}}
{G_{ii}}\sin\int^{r}\frac{\omega^{2}G_{ii}}{L},\qquad
r_{tp}<r\ll r_{s}
\end{equation}
We should note that the solutions in three regions are applied for both transversal and longitudinal equations.\\
In the next step, we consider the cases that three regions overlap.
As we mentioned before, regions 2 and 3 overlap close to the
horizon. By expanding the solutions (79) for large $\omega$'s near
the horizon we have following equation,
\begin{eqnarray}
\phi_{wkb}\simeq (4\pi T_{s})^{\frac{1}{2}}(r_{s}-r)^{\frac{1}{2}}\left\{C_{1}\cos
\left[\upsilon-\tilde{\omega}\log(r_{s}-r)\right]+
C_{2}\sin\left[\upsilon-\tilde{\omega}\log(r_{s}-r)\right]\right\},\quad r\rightarrow r_{s}
\end{eqnarray}
where $\upsilon=\int_{0}^{r_{1}} \frac{\omega^{2}G_{ii}}{L}$. By comparing relations (76) and (80) we find that,
\begin{equation}
C_{1}=-iC_{2}=\frac{C_{h}}{(4\pi T_{s})^{\frac{1}{2}}}e^{-i\upsilon}.
\end{equation}
Next we consider the near boundary region $r\ll r_{s}$. We know that
for the large $\omega$'s, the UV region overlaps with the WKB
region. For matching the UV solutions (72) and (73) for large
$\omega$'s, we need the following expansion for Bessel functions,
\begin{eqnarray}
J_{\nu}(x) &\simeq&\sqrt{ \frac{2}{\pi x}}\left[\cos
(x-\frac{\nu\pi}{2}-\frac{\pi}{4})+...\right],
\qquad\qquad |x|\rightarrow\infty \nonumber\\
Y_{\nu}(x) &\simeq&\sqrt{ \frac{2}{\pi x}}\left[\sin
(x-\frac{\nu\pi}{2}-\frac{\pi}{4})+...\right] .\qquad\qquad
|x|\rightarrow\infty
\end{eqnarray}
So, the large $\omega$'s expansion for solutions (72) and (73)
become,
\begin{eqnarray}
\phi_{uv}&=&A_{1}\sqrt{\frac{2}{\pi\gamma \omega}}\left[\cos \left(\gamma \omega r-
\frac{\pi}{2}(\frac{2d-\theta}{d})\right)\right]+ \nonumber\\
&&A_{2}\sqrt{\frac{2}{\pi\gamma \omega}}\left[\sin \left(\gamma \omega r-
\frac{\pi}{2}(\frac{2d-\theta}{d})\right)\right],\;\qquad\qquad z=1
\end{eqnarray}
\begin{eqnarray}
\phi_{uv}&=&A_{1}\sqrt{\frac{2z}{\pi\omega r^{z-1}}}\left[\cos \left(\frac{\omega r^{z}}{z}
-\frac{\pi}{2}(\frac{(z+1)d-\theta}{zd})\right)\right]+ \nonumber\\&&
A_{2}\sqrt{\frac{2z}{\pi\omega r^{z-1}}}\left[\sin\left(\frac{\omega r^{z}}{z}
 -\frac{\pi}{2}(\frac{(z+1)d-\theta}{zd})\right)\right].\qquad\qquad  z>1
\end{eqnarray}
In the other hand for WKB solutions in the $r\ll r_{s}$ we have,
\begin{eqnarray}
\phi_{wkb}&\simeq&\frac{C_{1}}{\sqrt{\gamma}}\cos(\gamma \omega r)+
\frac{C_{2}}{\sqrt{\gamma}}\sin(\gamma \omega r),\quad r\ll r_{s}\qquad\mathrm{for}\quad z=1 \nonumber\\
\phi_{wkb}&\simeq&\frac{C_{1}}{\sqrt{r^{z-1}}}\cos( \frac{\omega r^{z}}{z})+
\frac{C_{2}}{\sqrt{r^{z-1}}}\sin( \frac{\omega r^{z}}{z}),\quad r\ll r_{s}\qquad\mathrm{for}\quad z>1
\end{eqnarray}
By comparing between equations (83), (85) for $z=1$ and (84), (85)
for $z>1$ give us following equation,
\begin{eqnarray}
C_{1}&=&A_{1}e^{-i\frac{\pi}{2}(\frac{2d-\theta}{d})}\sqrt{\frac{2}{\pi \omega}},
\qquad A_{2}=iA_{1} \qquad\qquad\mathrm{ for}\quad z=1 \nonumber\\
C_{1}&=&A_{1}e^{-i\frac{\pi}{2}(\frac{(z+1)d-\theta}{zd})}\sqrt{\frac{2z}{\pi \omega}}
,\qquad A_{2}=iA_{1} \qquad\qquad \mathrm{for}\quad z>1
\end{eqnarray}
Finally, all coefficients depend on determination of $A_{2}$. By imposing unit normalization of the function
 $\psi=\frac{1}{\sqrt{\mathcal{L}}}\phi$ at $r = r_{b}$, i.e. the point where the string is attached,
one can  find this coefficient.
\paragraph{Infinite Quark Mass} In this case the endpoint of string is attached to the boundary
$r_{b} = r_{F}\rightarrow0$ and we normalize the wave-functions on
this location. Then by imposing the
$\psi(r_{F}\rightarrow0,\omega)=1$ we have,
\begin{eqnarray}
A_{2}&=&\frac{-\pi R(\frac{\omega}{2})^{\frac{3d-2\theta}{2d}}\gamma^{1-\frac{\theta}{d}}}
{r_{F}^{\frac{\theta}{d}}\Gamma\left[\frac{3d-2\theta}{2d}\right]},
\qquad\qquad z=1\nonumber \\
A_{2}&=&\frac{-\pi R(\frac{\omega}{2z})^{\frac{(z+2)d-2\theta}{2zd}}}{r_{F}^{\frac{\theta}
{d}}\Gamma\left[\frac{(z+2)d-2\theta}{2zd}\right]}
.\qquad\qquad z>1
\end{eqnarray}
Consequently, by using the equations (86) and (87) with (81) we take
following result,
\begin{eqnarray}
C_{h}&=&\frac{\sqrt{\pi} (\frac{\omega\gamma}{2})^{1-\frac{\theta}{d}}(4\pi T_{s})^{\frac{1}{2}}
 R}{r_{F}^{\frac{\theta}{d}}\Gamma\left[\frac{3d-2\theta}{2d}\right]}e^{i(\upsilon-\frac{\pi}{2}(\frac{d-\theta}{d}))},
\qquad\qquad z=1 \nonumber\\
C_{h}&=&\frac{\sqrt{\pi} (\frac{\omega}{2z})^{\frac{d-\theta}{zd}}(4\pi T_{s})^{\frac{1}{2}}R}
{r_{F}^{\frac{\theta}{d}}\Gamma\left[\frac{(z+2)d-2\theta}{2zd}\right]}
e^{i(\upsilon-\frac{\pi}{2}(\frac{d-\theta}{zd}))}.\qquad\qquad z>1
\end{eqnarray}
Eventually from the above expressions we derive the coefficient
$\psi_{h}$ as,
\begin{eqnarray}
\psi_{h}&=&\frac{\sqrt{\pi} (\frac{\omega\gamma}{2})^{1-\frac{\theta}{d}} }{r_{s}^{\frac{\theta}{d}-1}
\Gamma\left[\frac{3d-2\theta}{2d}\right]}e^{i(\upsilon-\frac{\pi}{2}(\frac{d-\theta}{d}))}\left\{
\begin{array}{ll}
1  &  \perp\\
\sqrt{\frac{G_{ii}(r_{s})}{N(r_{s})}}& \parallel\\
\end{array}
\right.,\qquad\qquad z=1\\
\psi_{h}&=&\frac{\sqrt{\pi} (\frac{\omega}{2z})^{\frac{d-\theta}{zd}}}{r_{s}^{\frac{\theta}{d}-1}
\Gamma\left[\frac{(z+2)d
-2\theta}{2zd}\right]}
e^{i(\upsilon-\frac{\pi}{2}(\frac{d-\theta}{zd}))}\left\{
\begin{array}{ll}
1  &  \perp\\
\sqrt{\frac{G_{ii}(r_{s})}{N(r_{s})}}& \parallel\\
\end{array}
\right..\qquad\qquad z>1
\end{eqnarray}
 By inserting these expressions in equation (66) we get,
 \begin{eqnarray}
\mathrm{Im }G_{R}^{\perp}&=& \frac{R^{2}\omega}{2\acute{\alpha}}
\frac{(\frac{\omega\gamma}{2})^{2-\frac{2\theta}{d}} }{r_{F}^{2\frac{\theta}{d}}\left[\Gamma\left(\frac{3d-2\theta}{2d}\right)\right]^{2}},
\qquad\qquad z=1 \nonumber\\
\mathrm{Im}G_{R}^{\perp}&=&\frac{ R^{2}\omega}{2\acute{\alpha}}
\frac{ (\frac{\omega}{2z})^{\frac{2(d-\theta)}{zd}}}{r_{F}^{2\frac{\theta}{d}}
\left[\Gamma\left(\frac{(z+2)d-2\theta}{2zd}\right)\right]^{2}} \qquad\qquad z>1
\end{eqnarray}
The longitudinal component of retarded correlator can be found
easily by the relations (65) and (66). As mentioned before the
imaginary part of retarded correlator is proportional to the
conserved current. By using  the above expression for imaginary part
of retarded correlator and equation (68), we determine the spectral
densities associated to the Langevin dynamics in the limit
$\omega\gg\frac{1}{r_{s}}$,
  \begin{eqnarray}
\rho_{\perp}(\omega)&\simeq &\gamma^{-2}\rho_{\parallel}(\omega),\qquad\qquad z=1 \nonumber\\
\rho_{\perp}(\omega)&\simeq & \rho_{\parallel}(\omega).
\qquad\qquad\quad\: z>1
\end{eqnarray}
So this  result is interesting only for $z=1$ for the transversal
and the longitudinal components of spectral density. In the region
$z>1$, this situation approximately converts to the equality for
components of spectral densities.
\paragraph{Finite Quark Mass} Here we are going to study  the finite mass quark which is  a similar work
for an infinitely case. Now, also  we use the relations
 (72) and (73) but with the normalization condition at the cutoff $r_{b}=r_{q}\neq0$. In this case one can obtain,
 \begin{eqnarray}
A_{2}&=&\sqrt{ \frac{L(r_{q})}{r_{q}}}\left[-iJ_{\frac{3d-2\theta}{2d}}(r_{q}\gamma \omega)
+Y_{\frac{3d-2\theta}{2d}}(r_{q}\gamma \omega)\right],\qquad\qquad z=1 \nonumber\\
A_{2}&=& \sqrt{\frac{L(r_{q})}{r_{q}}}\left[-iJ_{\frac{(z+2)d-2\theta}{2zd}}(\frac{r_{q}^{z}\omega}{z})+Y_{\frac{(z+2)d-2\theta}{2zd}}
(\frac{r_{q}^{z}\omega}{z})
\right].\qquad\qquad z>1
\end{eqnarray}
The other coefficients can be derived trough the above relation as with the previous way for the infinite mass case.
\subsection{Langevin Diffusion constants via the retarded correlator}
So far, we found the correlators and spectral densities which are required to
 establish the generalized Langevin equation. Now we want to find the diffusion coefficients
 from the information of last section for both $z=1$ and $z>1$ case.
We consider a long-time limit which makes the generalized Langevin
equation. This limit is expressed in the the zero-frequency limit of
the Green's functions [22-23] and [47].
 Therefore, we investigate the zero-frequency limit of the Green's functions which
  allow us to give the analytic results for the diffusion constants.
 The diffusion constant is defined in terms of the symmetric correlator $G_{sym}$ as,
 \begin{equation}
 \kappa=\lim_{\omega\rightarrow0}G_{sym}=-2T_{s}\lim_{\omega\rightarrow0}\frac{G_{R}(\omega)}{\omega}
 \end{equation}
By going back to the definition of correlator (64), it seems the evaluation of
wave-function in the zero-frequency limit is necessary. For this purpose,
 we write the small frequency limit for the horizon asymptotic of the $\psi _{R}$'s in (46),
\begin{equation}
\psi_{R}(r,\omega)=\psi_{h}(r_{s}-r)^{-i\frac{\omega}{4\pi T_{s}}}\simeq(1-\frac{-i\omega}{4\pi T_{s}}\log\mid r_{s}-r\mid+...)
\end{equation}
This solution reduces to the $\psi_{R} =\psi_{h}$ in the strict
$\omega= 0 $ limit.  It Matches with the boundary solution results
$\psi_{h} = 1 $ both for transverse and longitudinal modes. This
condition is applied consistently for  finite and infinite massive
quarks, since the radius value for boundary does not effect in (64)
for low frequency limit.
 Therefore, by using the explicit expressions (66) in equation (94) and $\psi_{h} = 1$, we receive
 the following results,
\begin{eqnarray}
 \kappa_{\perp} =\frac{R^{2}r_{s}^{\frac{2\theta}{d}-2}}{\pi\acute{\alpha}
 r_{F}^{\frac{2\theta}{d}}}T_{s}\qquad\qquad,\qquad\qquad
 \kappa_{\parallel}= \frac{16 \pi R^{2}r_{s}^{\frac{2\theta}{d}+2z-4} T_{s}^{3}}
 {\acute{\alpha}r_{F}^{\frac{2\theta}{d}}h^{2}(r_{s})\left(\frac{h'(r_{s})}{h(r_{s})}+
\frac{-2(z-1)}{r_{s}}\right)^{2}}
 \end{eqnarray}
From the above expression for the diffusion constants, it is obvious that there is
not any dependence on dynamical exponent $z$ for transversal component but for
longitudinal component there is so. The ratio between
 transversal and longitudinal component can be written as,
\begin{equation}
\frac{\kappa_{\parallel}}{\kappa_{\perp}}=\left[\frac{4\pi r_{s}^{z}T_{s}}{2(1-z)+(2+d-z-\theta)
(v^{2}-r_{s}^{-2(z-1)})}\right]^{2}.
 \end{equation}
For the special case $z=1$ and using the definition of $T_{s}$ in
(18), we get,
\begin{equation}
\frac{\kappa_{\parallel}}{\kappa_{\perp}}=1+\frac{4(d-\theta)}{d(d+1-\theta)}\left[\frac{v^{2}}{1-v^{2}}\right]
\end{equation}
For particular gauge/gravity dualities the inequality
$\kappa_{\parallel}> \kappa_{\perp}$ has been noticed to hold [49].
In the absence of hyperscaling parameter $\theta$ one can check that
this inequality is maintained, but in the presence of $\theta$, it
seems that we need the $\theta<d$ condition. We not that at
condition NEC, the range $\theta> d $ is allowed. But, in range of
$\theta$ we have some instabilities in the gravity side. For the
range $z>1$ the ratio between the transversal and longitudinal
components is given by,
   \begin{equation}
\frac{\kappa_{\parallel}}{\kappa_{\perp}}=1+\frac{4(d-\theta)v^{2}}{d\left[(d+z-\theta)
r_{s}^{-2(z-1)}-(2+d-z-\theta)v^{2}\right]}
\end{equation}
It is obvious that the range $ \theta<d$ is also necessary for the
case $z>1$. In this case there is also another condition,
$r_{s}^{2(z-1)}<\frac{d+z-\theta}{v^{2}(d+2-z-\theta)}$, with
$r_{s}$ defined in (11). In the special case $\theta=d$ the
universal inequality $\kappa_{\parallel}> \kappa_{\perp}$ convert to
the equality $\kappa_{\parallel}=\kappa_{\perp}$ for both $z=1$ and
$z>1$. For this special $\theta$ in the last section,
we found that the Hawking temperature and modified temperature are identical.\\
The jet-quenching parameters can be defined in terms of the
diffusion constants as [22-23],
\begin{equation}
\hat{q}^{\perp}=\frac{2\kappa^{\perp}}{v},\qquad\qquad \hat{q}^{\parallel}=\frac{\kappa^{\parallel}}{v}
\end{equation}
Therefore we obtain,
\begin{eqnarray}
 \hat{q}^{\perp} &=& 2\frac{R^{2}r_{s}^{\frac{2\theta}{d}-2}}{v\pi\acute{\alpha}r_{F}^{\frac{2\theta}{d}}}T_{s} \\
 \hat{q}^{\parallel} &=& \frac{ \kappa_{\perp}}{v}
\left[\frac{4\pi r_{s}^{z}T_{s}}{2(1-z)+(2+d-z-\theta)(v^{2}-r_{s}^{-2(z-1)})}\right]^{2}
 \end{eqnarray}
\subsection{The diffusion constants via the membrane paradigm }
This method allows us to achieve the diffusion constants directly
from the action (26), in here we do not  need to derive the
wave-function as a pervious section, which is described in detail by
[50]. By using this method
 for the following  metric background,
\begin{equation}
ds^{2}=a(r)^{2}dt^{2}+b(r)^{2}dr^{2}+c_{i}^{2}(r)dx_{i}^{2},
\end{equation}
we obtain the transversal and longitudinal diffusion constant as,
\begin{eqnarray}
 \kappa_{\perp} &=& \frac{c_{i}^{2}(r_{s})}{\pi\acute{\alpha}}T_{s}, \nonumber\\
 \kappa_{\parallel} &=& \pm\frac{16\pi}{\acute{\alpha}} \frac{|a^{2}|b^{2}}{c_{i}^{2}
 \left(\frac{a^{2}}{c_{i}^{2}}\right)' \left|\left(\frac{a^{2}}{c_{i}^{2}}\right)'
 \right| } \left|_{r=r_{s}}\right. T_{s}^{3}.
 \end{eqnarray}
 By inserting the component of metric background (2),  we receive to the expected result (96) for diffusion constants.
 In the next section we will study $R$-charged black hole  with hyperscaling violation.
\section{$R$-charged black holes with hyperscaling violation}
Generally, the $R$-charged black hole have three independent charges
and are static solutions of $N=2$ supergravity. The bosonic part of
the effective gauged supersymmetric $N=2$ Lagrangian  describes the
coupling of vector multiples in  supergravity [51] which is given by
the following expression,
\begin{equation}
e^{-1}L=\left(\frac{\Re}{2}-\frac{1}{2}g_{xy}\partial_{\mu}\phi^{x}\partial^{\mu}\phi^{y}-
\frac{1}{4}a_{IJ} F_{\mu\nu}^{I}F^{\mu\nu
J}-g^{2}V+\frac{e^{-1}}{48}\epsilon^{\mu\nu\rho\sigma\lambda}C_{IJK} F_{\mu\nu}^{I}
F_{\rho\sigma}^{J}A_{\lambda}^{K}\right),
\end{equation}
where $e=\sqrt{-g}$ is the determinant of veilbein, $\Re\:$ is the Ricci scalar,
$g_{xy}  $ is a metric on the scalar manifold,
 $a_{IJ}$ is a kinetic gauge coupling of the field strength ,  $g$ is a constant gauge coupling, $g_{xy},
 a_{IJ}$ and $V$ are functions of scalar field $\phi^{x}  $.
 The variation of Lagrangian (105) with respect to $g_{xy}$ ,
 $\phi^{x}$  and  $F_{\mu\nu}^{I}$ gives the field equations of motion.
 In Ref [28] we found the deformed R-charged black hole metric background with hyperscaling violation.
 In this section we use the form of R-charged black hole
 with hyperscaling violation in the flat space $k=0$.
 The metric background of this black hole is given by,
\begin{equation}
ds^{2}=(\frac{r}{r_{F}})^{2\frac{\theta}{3}}\frac{R^{2}}{r^{2}}\left[H^{\frac{1}{3}}(-hdt^{2}+
dx_{i}^{2})+H^{-\frac{2}{3}}h^{-1}dr^{2}\right],
\end{equation}
with
\begin{eqnarray}
H&=&(\frac{r}{r_{F}})^{\theta}+Q_{1}r^{2}+Q_{2}r_{F}^{\theta}
r^{4-\theta}+Q_{3}r_{F}^{2\theta}r^{6-2\theta},\nonumber\\
h&=&1-(\frac{r}{R})^{2}(\frac{r}{r_{0}})^{2-\theta}H^{-1},
\end{eqnarray}
where $Q_{1}=\frac{q_{1}+q_{2}+q_{3}}{R^{2}}$,
$Q_{2}=\frac{q_{1}q_{2}+q_{2}q_{3}+q_{1}q_{3}}{R^{4}}$, and
$Q_{3}=\frac{q_{1}q_{2}q_{3}}{R^{6}}$. Notice that in the definition
of metric background (106), we use the radial coordinate
transformation $r\rightarrow\frac{R^{2}}{r}$. We introduce parameter
$r_{h}$ as the location of horizon in the $r$ coordinate such that
it is the largest root of $h=0$.
 Now, by using  the above information, we are ready to proceed as pervious section.
\subsection{Trailing string and drag force}
In order to study the stochastic motion of the quark in a plasma on the boundary of R-charged black hole
 with hyperscaling violation, we repeat the process of section 2 in this section. The trailing string
 corresponding to a quark moving on the boundary of R-charged black hole with a constant velocity $v$ -through the parametrization (6)-
 is characterized by the following induced world-sheet metric,
 \begin{equation}
 g_{ab}= (\frac{R}{r})^{2}(\frac{r}{r_{F}})^{\frac{2\theta}{d}} H(r)^{\frac{1}{3}}  \left(
 \begin{array}{ll}
 v^{2}-h(r)  & \hbox{$ v\xi'(r)$} \\
 v\xi'(r)   & \hbox{$\left(h(r)H(r)\right)^{-1}+\xi'^{2}$}
 \end{array}
 \right).
 \end{equation}
 Constructing the Nambu-Goto action we find the momentum $\pi_{\xi}$ flowing from the bulk to the horizon which is equal to the drag force,
 \begin{equation}
F_{drag}= \pi_{\xi}=\frac{v R^{2}H^{\frac{2}{3}}(r_{s})}{2\pi \acute{\alpha}r_{F}^{2}}
\end{equation}
 The stretched horizon $r_{s}$  is defined through the relation $h(r_{s})=v^{2}$ and this expression reduces to solve the following equation,
\begin{equation}
(1-v^{2})r_{s}^{\theta}(1+ax+bx^{2}+cx^{3})=0,\qquad\qquad\mathrm{with}\quad \textit{x}=\textit{r}_{s}^{2-\theta}
\end{equation}
where $a=Q_{1}r_{F}^{\theta}$, $b=Q_{2}r_{F}^{2\theta}+\frac{r_{F}^{\theta}}{R^{2}r_{0}^{2-\theta}(1-v^{2})}$ and $c=Q_{3}r_{F}^{3\theta}$.
The stretched horizon for $\theta\neq2$ in the terms of $a$, $b$ and $c$ is given by,
\begin{eqnarray}
x&=&\frac{1}{3}\left[\frac{B}{2c}-2\frac{(3ac-b^{2})}{cB}-\frac{b}{c} \right]\nonumber\\
 \mathrm{with}\quad B&=&\left[12\sqrt{3}\sqrt{4a^{3}c-a^{2}b^{2}-18abc+4b^{3}+27c^{2}}c+36abc-8b^{3}-108c^{2}\right]^{\frac{1}{3}}
\end{eqnarray}
For the special case $\theta=2$, in order to receive $r_{s}\neq0$ there must be the condition $1+a+b+c=0$.
If we diagonalize the world-sheet induced metric (108), then the modified temperature is obtained as,
\begin{equation}
T_{s}^{2}=\frac{h'(r_{s})h(r_{s})H(r_{s})}{16\pi^{2}}\left[\frac{4(\frac{\theta}{3}-1)}{r_{s}}+\frac{h'(r_{s})}{h(r_{s})}
+\frac{2}{3}\frac{H'(r_{s})}{H(r_{s})}\right]
\end{equation}
By calcuting $h'(r_{s})$ in detail one can obtain that for
$\theta=3-1=2$  this expression becomes zero and hence the
temperature. For this value of $\theta$, we obtained in Ref.
\cite{P28} that the temperature and total particle number got zero.
Therefore for $\theta=2$ the Hawking temperature and the modified
temperature both become equal to zero.
\subsection{Fluctuations of trailing string}
In order to investigate the fluctuations of trailing string we utilize the quadratic Nambu-Goto action (26) with
\begin{equation}
 N(r)= \frac{h(r)(\frac{R}{r})^{4}(\frac{r}{r_{F}})^{\frac{4\theta}{d}}H(r)^{\frac{2}{3}}-C^{2}}
 {(\frac{R}{r})^{2}(\frac{r}{r_{F}})^{\frac{2\theta}{d}}H(r)^{\frac{1}{3}}(h(r)-v^{2})},\qquad\qquad G_{ii}=(\frac{R}{r})^{4}
 (\frac{r}{r_{F}})^{\frac{4\theta}{d}}H(r)^{\frac{1}{3}}
\end{equation}
Therefore, by using the above expressions the equations of motion (28) become,
\begin{eqnarray}
 &&\partial_{r}\left[\sqrt{H(r)\left(C^{2}-(\frac{R}{r})^{4}
(\frac{r}{r_{F}})^{\frac{4\theta}{d}}H(r)^{\frac{2}{3}}h(r)\right)\left(v^{2}-h(r)\right)
}\partial_{r}\delta X^{\perp}\right]\nonumber\\
&&+\frac{\omega^{2}(\frac{R}{r})^{4}
 (\frac{r}{r_{F}})^{\frac{4\theta}{d}}}{\sqrt{H(r)\left(C^{2}-
(\frac{R}{r})^{4}
 (\frac{r}{r_{F}})^{\frac{4\theta}{d}} H(r)^{\frac{2}{3}}h(r)\right)\left(v^{2}-h(r)\right)}}\delta X^{\perp}=0
\end{eqnarray}
\begin{eqnarray}
 &&\partial_{r}\left[\frac{\left(C^{2}-(\frac{R}{r})^{4}
 (\frac{r}{r_{F}})^{\frac{4\theta}{d}}h(r)\right)H(r)^{\frac{3}{2}}}{\left(v^{2}-h(r)\right)^{\frac{1}{2}}
H(r)^{\frac{1}{6}}}\partial_{r}\delta X^{\parallel}\right]\nonumber\\
&&+\frac{\omega^{2}\left(C^{2}-
(\frac{R}{r})^{4}(\frac{r}{r_{F}})^{\frac{4\theta}{d}}H(r)^{\frac{2}{3}}h(r)\right)^{\frac{1}{2}}}
{\left(v^{2}-h(r)\right)^{\frac{3}{2}}H(r)^{\frac{1}{2}}}\delta X^{\parallel}=0.
\end{eqnarray}
At the world-sheet horizon $r\rightarrow r_{s}$  both of the above equations reduce to the equation (34) with $T_{s}$
 defined in (112). Therefore, the solution to this equation is similar to what obtained in relation (35).
  Near the boundary limit, $r\rightarrow r_{F}$,
of the equations (114) and (115), we receive to the relation (36) with $d=3$, since in this limit $h(r)$ and
$H(r)$ have the following behavior,
\begin{equation}
\lim_{r\rightarrow r_{F}\rightarrow0} H(r)=1,\qquad\qquad \lim_{r\rightarrow r_{F}\rightarrow0} h(r)=1 \,.
\end{equation}
So the solutions to the equations of motion near the boundary are similar as before, i.e relation (37) with $d=3$.
\subsection{Momentum correlator of trailing string}
The classical solutions that we obtained for a trailing string in the black hole background (106) are implemented
in computing the Langevin correlators. As we discussed before, these correlators involve two parts: real and imaginary.
\subsubsection{Real part of retarded correlator}
The real part of the correlators from the fluctuation modes (37) are obtained trough the relation (32) as,
\begin{equation}
Re\,G_{R}^{\perp}\simeq \gamma^{-2} Re\,G_{R}^{\parallel}\simeq \frac{\gamma\omega^{2}}{2\pi\acute{\alpha}(1-\frac{2\theta}{3})r_{F}}\left[
1+\frac{\gamma^{2}\omega^{2}r_{F}^{2}}{2(1-\frac{2\theta}{3})}+O(r_{F}^{4})\right].
\end{equation}
As this relation demonstrates, in the limit of  $r_{F}\rightarrow0$ there are some divergencies of the order of
$(\frac{1}{r_{F}})$. These divergencies can be canceled in a similar way to the section 2 by the identification
 of $\Delta M(r_{F})=-\frac{R^{2}}{2\pi\acute{\alpha}(1-\frac{2\theta}{3}) r_{F}}$  in the boundary counterterm (56).
\subsubsection{Imaginary part of retarded correlator}
For computing the imaginary part of retarded correlator, we require relation (64). The expressions $g_{\perp}^{rr}$
and $g_{\parallel}^{rr}$  in relation (31) reduce to the following for metric background (106),
\begin{eqnarray}
&&\emph{g} ^{rr}_{\perp}\simeq 4\pi T_{s}(r_{s}-r)G_{ii}(r_{s}),\qquad \emph{g} ^{rr}_{\parallel}
\simeq 4\pi T_{s}(r_{s}-r)N(r_{s}),\quad
\quad r\rightarrow r_{s}, \nonumber\\
&&\mathrm{with}\qquad \textit{N}(r_{s})=\frac{16 \pi^{2}R^{2}T_{s}^{2}}{r_{F}^{\frac{2\theta}{d}}H^{\frac{2}{3}}
(r_{s})h'^{2}(r_{s})}r_{s}^{\frac{2\theta}{d}-2}
\end{eqnarray}
By inserting the above expressions in the (64) and using the solution (45) for $\psi$ we write,
\begin{equation}
\mathrm{Im} \textit{G}_{R}^{\perp}=-\frac{\textit{G}_{ii}(r_{s})\mid \psi_{h}^{\perp}\mid^{2}}{2\pi
 \acute{\alpha}}\omega,\qquad\qquad\mathrm{Im} \textit{G}_{R}^{\parallel}=-\frac{\textit{N}(r_{s})
 \mid \psi_{h}^{\parallel}\mid^{2}}{2\pi \acute{\alpha}}\omega\,.
\end{equation}
The spectral densities form these correlators can be found via the relation (68).
To obtain the high-frequency behavior of the Langevin correlators, we will implement the WKB method in the next subsection.
\subsection{The WKB approximation at large frequency}
In this section we implement the WKB method to obtain the high-frequency behavior of spectral densities
. Everything is the same as section 2, except that the function $L$ has the following definition in
the Schr\"{o}dinger potential (69),
\begin{equation}
L=\sqrt{H(r)\left(C^{2}-(\frac{R}{r})^{4}(\frac{r}{r_{F}})^{\frac{4\theta}{d}}h(r)H^{\frac{2}{3}}(r)\right)\left(v^{2}-h(r)\right)}
\end{equation}
As before, we divide the range of  $r_{F}<r<r_{H}$  in three region and derive the solutions to the Schr\"{o}dinger equations.
\paragraph {Near Boundary}: $r\ll r_{s}$\\
In this region with,
\begin{eqnarray}
L =\frac{R^{2}r^{\frac{2\theta}{3}-2}}{r_{F}^{\frac{2\theta}{3}}\gamma}\quad,\quad\quad
    N=\frac{R^{2}r^{\frac{2\theta}{3}-2}}{r_{F}^{\frac{2\theta}{3}}}
  \end{eqnarray}
and so, the Schr\"{o}dinger potential,
\begin{eqnarray}
V_{s}\simeq  -\gamma^{2}\omega^{2}+\frac{(\frac{\theta}{3}-1)(\frac{\theta}{3}-2)}{r^{2}}
\end{eqnarray}
we receive to the following solutions,
\begin{equation}
\phi=A_{1}\sqrt{r}J_{\frac{3}{2}-\frac{\theta}{3}}(r\gamma \omega)+A_{2}\sqrt{r}Y_{\frac{3}{2}-\frac{\theta}{3}}(r\gamma \omega),
\end{equation}
\paragraph{Near Horizon}: $r\simeq r_{s}$ \\
In this region, one can obtain the following for $L$ function,
\begin{equation}
L=(4\pi T_{s})\frac{R^{2}r_{s}^{\frac{2\theta}{2}-2}}{r_{F}^{\frac{2\theta}{3}}} H^{\frac{1}{3}}(r_{s})(r_{s}-r).
\end{equation}
If we implement the above relation and the relation (119) for $N(r_{s})$ in the schr\"{o}dinger potential then we get,
\begin{equation}
V_{s}\simeq -(\tilde{\omega}^{2}+\frac{1}{4})\frac{1}{(r-r_{s})^{2}}\qquad\qquad r\rightarrow r_{s}
\end{equation}
where $ \tilde{\omega}=\frac{\omega}{4\pi T_{s}}$. The solution to the Schr\"{o}dinger equation (69)
 after substituting this potential is identical to what found in solution (76).
\paragraph{WKB region}: $r_{tp}<r\ll r_{s}$\\
In this region, for large $\omega$'s, the first term in equation (69) dominates and the Schr\"{o}dinger potential becomes,
\begin{equation}
V_{s}\simeq -\frac{\omega^{2}G_{ii}^{2}}{R^{2}},\qquad\qquad r_{tp}<r\ll r_{s}.
\end{equation}
The turning point for large $\omega$s is found as,
\begin{eqnarray}
r_{tp}= \frac{ \sqrt{(\frac{\theta}{3}-1)(\frac{\theta}{3}-2)}}{\gamma \omega}
\end{eqnarray}
We should note that the solutions in three regions are applied for both transversal and longitudinal equations.\\
Now, with the above information, we are able to find the coefficients by considering the limits that three regions overlap. Eventually,
By repeating the process followed in section 2, we obtain the following expression for $\psi_{h}$ for the infinite massive quark case,
\begin{eqnarray}
\psi_{h}&=&\frac{\sqrt{\pi} (\frac{\omega\gamma}{2})^{1-\frac{\theta}{3}} }{r_{s}^{\frac{\theta}{3}-1}
\Gamma\left[\frac{3}{2}-\frac{\theta}{3}\right]}e^{i(\upsilon-\frac{\pi}{2}(\frac{3-\theta}{3}))}\left\{
       \begin{array}{ll}
           1  &  \perp\\
                  \sqrt{\frac{G_{ii}(r_{s})}{N(r_{s})}}& \parallel\\
                        \end{array}
                        \right.
\end{eqnarray}
and then by inserting the above expressions in equation (64) we attain,
 \begin{eqnarray}
\mathrm{Im }G_{R}^{\perp}&=& \frac{R^{2}\omega}{2\acute{\alpha}}\frac{(\frac{\omega\gamma}{2})^{2-\frac{2\theta}{3}} }
{r_{F}^{2\frac{\theta}{3}}\left[\Gamma\left(\frac{3}{2}-\frac{\theta}{3}\right)\right]^{2}}.
\end{eqnarray}
The longitudinal component can be found in a same way and by using the relation (66).
 The finite mass case is obtained trough the way
 described in section 2.
\subsection{Langevin diffusion constants}
The diffusion coefficients can be found either from the direct evaluation of the correlators or using
the membrane paradigm. Both of these method give rise to the following identical results for diffusion constants,
\begin{eqnarray}
 \kappa_{\perp} =\frac{R^{2}H^{\frac{1}{3}}(r_{s})r_{s}^{\frac{2\theta}{3}-2}}
 {\pi\acute{\alpha}r_{F}^{\frac{2\theta}{3}}}T_{s}\qquad\qquad,\qquad\qquad
 \kappa_{\parallel}= \frac{16 \pi R^{2}r_{s}^{\frac{2\theta}{3}-2} T_{s}^{3}}
 {\acute{\alpha}r_{F}^{\frac{2\theta}{3}}H^{\frac{2}{3}}(r_{s})h'^{2}(r_{s})}
 \end{eqnarray}
There is a problem in computing the longitudinal component for the special case $\theta=2$.
 In this case both $H(r_{s})$ and $h'(r_{s})$ tend
to the zero, so the longitudinal diffusion constant tends to the
infinity. It seems that this value for $\theta$ is unacceptable. By
studying the Ref. [28], we perceive that the null energy condition
impose some condition on $\theta$ as
\begin{eqnarray}
\theta &\leq &0.365 \nonumber\\
\theta &\geq &1.315.
\end{eqnarray}
In a similar way to the previous work and because we see some instabilities in the range $\theta \geq 1.315$,
we choose the range $\theta \leq 1.315$.
This case is in analogy with the case $\theta<d$ in the previous section.\\
 The ratio between longitudinal and transversal components of diffusion constants is,
 \begin{equation}
 \frac{\kappa_{\parallel}}{ \kappa_{\perp}} =1+2\frac{h(r_{s})}{h'(r_{s})}\left[\frac{\frac{2\theta}{3}-2}{r_{s}}+
 \frac{H'(r_{s})}{(r_{s})}\right]
 \end{equation}
 We expect the universal inequality $ \kappa_{\parallel}> \kappa_{\perp}$ to be hold. Therefore,
 to remain the inequality for the following relation,
\begin{equation}
 \frac{\kappa_{\parallel}}{ \kappa_{\perp}} =1+2\frac{v^{2}}{(1-v^{2})}\left[\frac{(\frac{r_{s}}{r_{F}})^{\theta}+
 \frac{2Q_{1}r_{s}^{2}}{3}+ \frac{Q_{2}r_{F}^{\theta}r_{s}^{4-\theta}}{3}}
 {2(\frac{r_{s}}{r_{F}})^{\theta}+Q_{1}r_{s}^{2}-Q_{3}r_{F}^{2\theta}r_{s}^{6-2\theta}}\right]
 \end{equation}
with $Q_{i}>0$, the condition $2+ax>cx^{3}$  must be satisfied. For $Q_{i}<0$, both of the numerator and denominator
 on the righthand
side of relation (133) can be negative or positive, where each of these situations put some conditions on $Q_{i}$ and $r_{s}$.
Since due to the relation (111), the $r_{s}$ is related to $Q_{i}$, $r_{F}$  and $v^{2}$ , to hold the inequality there should
 be some conditions on
$Q_{i}$, $r_{F}$ and $v^{2}$. The jet-quenching parameters can be obtained easily from the relation (100) as in the previous section.
\section{Summary}
In this paper, we  used AdS/CFT correspondence and studied the
stochastic motion of an external quark in a plasma. It corresponds
to  fundamental string whose end-point lies in the UV region of a
bulk black-hole background and is forced to move with velocity $v$.
By using the Nambu-Goto action, we obtained the equations of motion
for this string in the planar black holes with hyperscaling
violation background. The solution of  the corresponding equations
of motion lead us to find the classical profile of the trailing
string. Next,  we considered small fluctuations around the classical
string profile. These fluctuations satisfied the second-order radial
equations and related to the associated thermal correlators. We
achived the modified temperature $T_{s}$, it was felt by string
fluctuations. In order to have a positive temperature, we got a
constraint on $\theta$  for the case of ($z=1$). We derived the drag
force on the quark in the presence of dynamical exponent $z=1$ and
$z>1$. Here, we found that in the ultra-relativistic limit this
force becomes zero for the case $z=1$, while for the range $z>1$ is
not so. We computed holographically the full Langevin correlators
and the associated spectral densities, including real and imaginary
parts of correlators for large and small frequency limits. In the
large-frequency regime, the spectral densities are obtained via the
modified WKB method. We shown that only for the case $z=1$ the ratio
between transversal and longitudinal components of spectral density
depends on the velocity of quarks. For the range $z>1$ this relation
tends to a constant with the value of 1. We investigated the
large-time limit of the fluctuations and for the constant diffusion
we need small frequency modes to obtain the diffusion. In order to
hold inequality $\kappa_{\parallel}\geq\kappa_{\perp}$, we have
shown that the range of $\theta\leq d$ is an acceptable region for
$\theta$. This region is in agreement with what found in literature.
We note that the region $\theta> d$ leads to some instabilities on
the gravity side, so for this region we cannot have the physically
consistent theories. For the special case $\theta=d$ the inequality
$\kappa_{\parallel}\geq\kappa_{\perp}$ converts to the equality
$\kappa_{\parallel}=\kappa_{\perp}$ for both $z=1$ and $z>1$ cases.
In this case we found that the Hawking temperature and the modified
temperature become equal, however the world-sheet
horizon and the black hole horizon are different.\\
We repeated all of the above procedure for the R-charged black hole
with hyperscaling violation. For this black hole we found the range
$\theta\leq1.315$ can provide a plausible region for $\theta$ which
is in agreement with our results in Ref. [28]. We also realized that
the universal inequality of diffusion constants can be confirmed by
some constraints on the world-sheet horizon and charges.\\ As we
indicated before, we obtained the real part and imaginary part of
retarded correlator for both planer and R-charged black hole with
hyperscaling violation. For real part, we encountered some
divergencies in the limit $r\rightarrow r_{F}$. In that case we
introduced  a boundary counterterm action for the case $z=1$ in
planar black hole and R-charged black hole with hyperscaling
violation. For the case of $z>1$ in planar black hole with
hyperscaling violation, the boundary counterterm with the last
definition cannot help us to remove the divergencies. It seems that
a new definition of boundary counterterm is needed to overcome this
problem. We would like to investigate this problem in future works.
We are also interested to study about different  black holes with
hyperscaling violation and investigate the Langevin diffusion
process.

\end{document}